%% file: paper.tex
\documentclass[acmsmall]{acmart}
\AtBeginDocument{%
  \providecommand\BibTeX{{%
    Bib\TeX}}}
\newif\ifdraft\draftfalse{}
\newif\iflater\latertrue{}
\newif\ifaftersubmission\aftersubmissionfalse{}
\newif\ifsilly\sillytrue{}

\PassOptionsToPackage{names,dvipsnames}{xcolor}

\usepackage{pgfplots}
\usepackage{pgfplotstable}
\pgfplotsset{compat=1.17}
\usepackage{listings}
\usepackage{subcaption}
\usepackage{xspace}
\usepackage{wrapfig}
\include{macros}

\theoremstyle{definition}

\AtBeginDocument{%
  \providecommand\BibTeX{{%
    \normalfont B\kern-0.5em{\scshape i\kern-0.25em b}\kern-0.8em\TeX}}}

\setcopyright{cc}
\setcctype{by}
\acmDOI{10.1145/3798263}
\acmYear{2026}
\acmJournal{PACMPL}
\acmVolume{10}
\acmNumber{OOPSLA1}
\acmArticle{155}
\acmMonth{4}
\acmSubmissionID{oopslaa26main-p387-p}
\received{2025-10-10}
\received[accepted]{2026-02-17}

\begin{document}

\title{Fail Faster}
\subtitle{Staging and Fast Randomness for High-Performance PBT}

\author{Cynthia Richey}
\authornote{Both authors contributed equally to this work.}
\affiliation{%
  \institution{University of Pennsylvania}
  \city{Philadelphia}
  \state{Pennsylvania}
  \country{USA}
}
\email{lapwing@seas.upenn.edu}

\author{Joseph W. Cutler}
\authornotemark[1]
\affiliation{
  \institution{University of Pennsylvania}
  \city{Philadelphia}
  \state{Pennsylvania}
  \country{USA}
}
\email{jwc@seas.upenn.edu}

\author{Harrison Goldstein}
\affiliation{%
  \institution{University at Buffalo}
  \city{Buffalo}
  \state{New York}
  \country{USA}
}
\email{hgoldste@buffalo.edu}

\author{Benjamin C. Pierce}
\affiliation{%
 \institution{University of Pennsylvania}
 \city{Philadelphia}
 \state{Pennsylvania}
 \country{USA}}
\email{bcpierce@seas.upenn.edu}
\renewcommand{\shortauthors}{C. Richey, J. W. Cutler, H. Goldstein, and B. C. Pierce}

\begin{abstract}
Property-based testing (PBT) relies on generators for random test
cases, often constructed using embedded domain specific languages,
which provide expressive combinators for building and composing
generators.  The effectiveness of PBT depends critically on the speed
of these generators.
However, careful measurements show that the generator performance of
widely used PBT libraries falls well short of what is possible,
due principally to (1) the abstraction overhead of their
combinator-heavy style and (2) suboptimal sources of randomness.
We characterize, quantify, and address these bottlenecks.

To eliminate abstraction overheads, we propose a technique based on
multi-stage programming, dubbed \name.  We apply this technique to
leading generator libraries in OCaml and Scala 3, significantly
improving performance.
To quantify the performance impact of the randomness source, we carry
out a controlled experiment, replacing the randomness in the OCaml PBT
library with an optimized version.
Both interventions exactly preserve the semantics of generators,
enabling precise, pointwise comparisons.
Together, these improvements find bugs up to $13\times$ faster.
\end{abstract}




\maketitle

\section{Introduction}
\label{section:intro}

Property-based testing is a software testing technique that uses
random test inputs to validate logical specifications~\cite{quickcheck00}. A recent study
on PBT usage in industry~\cite{inpractice} shows that many
practitioners run their property-based tests very frequently, and with
a very short time budget.  The faster the tests fail, the better.

One important opportunity for performance improvement
is the \textit{generators} that produce random inputs to the properties under
test. These generators are written with the help of \textit{generator
libraries}, usually expressed as embedded domain-specific languages (eDSLs)
that provide combinators for building and composing generators.

However, careful
measurements show that the performance of existing generator libraries falls far short of what
is possible,
for two reasons.
First, the high-level design approach followed by generator combinator
libraries introduces layered abstractions  that are difficult for
compilers to optimize.
Second, the calls a generator makes to its \rand{} can constitute a large
proportion of its run time, magnifying the cost of an inefficient
implementation.
In this paper, we characterize, quantify, and address these issues.
% We address the overhead of
% many common generator abstractions using \textit{multi-stage
%   programming} (or {\em staging}), a well-studied technique for
% building fast DSLs “without regret”~\cite{lms}.  We refer to this approach as \name.
% %

%
We address the overhead of
many common generator abstractions using \textit{multi-stage
  programming} (or {\em staging}), a well-studied technique for
building fast DSLs "without regret"~\cite{lms}. 
The simple monadic structure of typical PBT generators lend themselves naturally
to well-known staging transformations. Beyond these basic techniques, we observe
that common usage patterns of fundemental PBT-specific combinators---e.g., weighted choices with statically known options---are amenable
to advanced staging techniques~\cite{partially-static,kovacs24, carrette05}.

% Beyond the standard
% staging transformations, we apply advanced 
% that let us leverage our domain-specific understanding of PBT generators to achieve the fastest possible code.

We refer to this approach as \name.
To demonstrate its generality, we implement staged generator DSLs in
two strict functional languages: OCaml and Scala 3.  In OCaml, we
build \camlname{}, based on \bq{}---a high-quality PBT library in
OCaml, authored by software engineers at the trading firm Jane Street~\cite{bq}.  In Scala 3, we build \scalaname{}, based on ScalaCheck---the
language's standard PBT library~\cite{sc}.

We isolate and quantify the performance benefits of faster randomness, showing
that a DSL's choice of \rand{} has a dramatic effect on its performance. Based
on the observation that the OCaml implementation of \bq{}'s \rand{} is slow in
a way that is simple to fix, we perform a controlled experiment, comparing fast and slow
versions of the library to analyze the impact of faster \rands{} on
generator performance. We perform this comparison in \camlname{}.

Both interventions preserve semantics on the nose: given the same
random seed, generators written using \camlname or \scalaname produce
exactly the same sequence of values as equivalent generators using \bq
or ScalaCheck.  This semantic equivalence enables pointwise---as
opposed to distributional---comparisons of bug-finding effectiveness,
ensuring that, if an \name{} generator finds bugs faster, this is
attributable \textit{solely} to the staging and \rand interventions,
not lucky choice of seeds.

We evaluate the optimizations with a series of case studies in each
generator library, and show that the \name{} generators run faster
than their unstaged equivalents. In \camlname, they run up to $7\times$
faster, and in \scalaname, up to $13\times$ faster. Using our improved version
of \bq's \rand in \camlname, we see even greater gains---more than $12\times$---showing that
both staging and fast randomness play distinct and complementary roles in improving generator performance.
Further,
we show that \camlname generators find bugs faster by running our case studies
in the Etna platform, which uses generated values as
inputs to a buggy system and measures how quickly different
generators can detect the bugs. In Etna,
\camlname generators find bugs up to $2.5\times$ faster on average than
their \bq counterparts. With fast randomness,
the average speedup rises to $3.8\times$,
with some cases exceeding $13\times$. Finally, we review related libraries
in other languages including Racket, F\#, Haskell, Rust, Clojure, and Swift,
and discuss how the \name{} technique could be applied to them.

In summary, we show that PBT generator DSLs incur significant
performance costs across languages and present two interventions that
significantly improve their efficiency while preserving their
idiomatic style.  Concretely, we offer the following contributions:
\begin{enumerate}
\item We identify two key sources of inefficiency in PBT generator
libraries---abstraction overhead and choice of \rand---both of which
significantly impact performance (Section~\ref{section:motiv}).
\item We present \name{}, a staging technique that eliminates the
abstraction overhead of generators. We apply \name{} to standard
generator libraries in both OCaml and Scala 3, showcasing its
generality (Section~\ref{section:faster-generators}).
% \item We present a novel methodology---enabled by the pointwise semantic equivalence of original and \name{}-optimized generators---for quantifying 
%       the precise speedup achieved using staging and fast randomness,
%       in which the performance benefit of each intervention can be 
%       attributed to the interventions themselves, as opposed to 
%       random chance (Section~\ref{section:eval}).
\item We demonstrate that writing generators using the \name{} technique combined with fast
\rands{} yields substantial performance improvements, both separately
and in combination, and we show that these performance improvements
extend to significantly improved bug-finding speed (Section~\ref{section:eval}).
\end{enumerate}

Section~\ref{section:other-langs} discusses how the \name{}
technique could be applied to PBT libraries in other
languages---Racket, F\#, Haskell, Rust, Clojure, and Swift.
Sections~\ref{section:related} and~\ref{section:concl} present related
and future work.

\section{What Are Generator Libraries, and Why Are They Slow?}
\label{section:motiv}
% \jwc{In this section we answer: ``What are monadic generator libraries, and why are they slow?''}
% \bcp{That's actually not a bad section title!}

% \subsection{Background: Monadic Generator DSLs}

Property-based testing~\cite{claessenQuickCheckLightweightTool2000} is an approach to software testing that centers around
executable specifications of programs called {\em properties}. For example, if a
programmer wants to test an invariant of a binary search tree (BST)
implementation they are working on, they may write a property like
\begin{center}
  \texttt{prop\_insert\_invariant t x = isBST t ==> isBST (insert x t)}
\end{center}
to check that for any tree \texttt{t} and value \texttt{x}, if \texttt{t} is
already a valid BST then inserting \texttt{x} into it also yields a valid BST.
Once a developer has a property, they test that property by executing
it on hundreds or thousands of random test inputs. These test inputs are usually
produced by a {\em generator}---a program written in some domain-specific
language (DSL) that allows the developer to express precisely how values should
be sampled.

\begin{figure}[h]
\begin{lstlisting}
module Bq : sig
  type 'a t
  val gen_int : int -> int -> int t
  val return : 'a -> 'a t
  val bind : 'a t -> ('a -> 'b t) -> 'b t

  val weighted_union : (int * 'a t) list -> 'a t

  val size : int t
  val with_size : int -> 'a t -> 'a t

  val fixed_point : ('a t -> 'a t) -> 'a t
  $\dots$
end
\end{lstlisting}
\caption{Some functions from the API of the \bq{} generator DSL.}
\label{fig:bq-api}
\end{figure}

The standard design for a generator DSL, introduced in Haskell's QuickCheck
library~\cite{quickcheck00} and copied in dozens of other frameworks,
is via an embedded {\em monadic} language~\cite{moggi91}. We use the syntax and types from OCaml's \bq library, which is presented in
Figure~\ref{fig:bq-api}, but similar DSLs can also be found in languages
like Haskell, Scala, Python, and many more.
The library provides some basic generators, for
example \texttt{gen\_int} for generating random integers in a range, along with
the functions \texttt{return} and \texttt{bind}.  The
generator \texttt{return x} is the constant generator, always generating the
value \texttt{x}. Running a generator \texttt{bind g k} runs the generator
\texttt{g}, producing a value \texttt{a}, and then runs the generator \texttt{k
a}.

Together, these three functions are the bare minimum for constructing arbitrary
random data generators: \texttt{gen\_int} provides a base source of randomness,
and \texttt{return} and \texttt{bind} allow generators to be composed to create
larger, more complex generators. Figure~\ref{fig:simple-bind} shows a generator
built with these operations; it first samples an int between $0$ and $100$,
names it $x$, samples another between $0$ and $x$, and then returns the pair of
them.

\begin{figure}[h]
  \begin{subfigure}{.49\textwidth}
\begin{lstlisting}
let int_pair : (int * int) Bq.t =
  Bq.bind (Bq.gen_int 0 100) (fun x ->
    Bq.bind (Bq.gen_int 0 x) (fun y ->
      Bq.return (x,y)))
\end{lstlisting}
\caption{A simple generator using \texttt{bind} explicitly.}\label{fig:simple-bind}
  \end{subfigure}
  \begin{subfigure}{.49\textwidth}
\begin{lstlisting}
let int_pair : (int * int) Bq.t =
  let%bind x = Bq.gen_int 0 100 in
  let%bind y = Bq.gen_int 0 x in
  return (x,y)
\end{lstlisting}
\caption{An equivalent generator using the macro for \texttt{bind}.}\label{fig:simple-macro}
  \end{subfigure}
  \caption{Simple monadic generators for a pair of ordered integers.}\label{fig:simple-pair}
\end{figure}

Most languages in which monadic APIs are common expose some sort of syntactic
sugar for them. In OCaml \footnote{OCaml actually has a few ways to implement
monadic syntax; this is the one provided by Jane Street's libraries.}, this
looks like \texttt{let\%bind x = e in e'} which desugars to
\texttt{bind e (fun x => e')}. Figure~\ref{fig:simple-macro} shows the same
generator, written in the monadic syntax.

\begin{wrapfigure}{r}{.49\textwidth}
  \vspace{1.5em}
\begin{lstlisting}[
    xleftmargin=0pt,
    xrightmargin=0pt,
    aboveskip=0pt,
    belowskip=0pt,
    basicstyle=\ttfamily\footnotesize,
    lineskip=0pt
  ]
let tree_of g = fixed_point (fun rg ->
  let%bind n = size in
  weighted_union [
    (1, return E);
    (n,
      let%bind x = g in
      let%bind l = with_size (n / 2) rg in
      let%bind r = with_size (n / 2) rg in
      return (Node (l,x,r)))
  ])
\end{lstlisting}
\caption{A generator using a variety of convenience functions.}\label{fig:tree_of}
\end{wrapfigure}

Generator libraries also include other functions that make generator
construction easier; some examples of these are included in
Figure~\ref{fig:bq-api}.  The \texttt{weighted\_union} function is a
particularly well-used one: it makes a weighted choice between different
generators, allowing the developer to combine different sub-generators into a
single program and tune the data distribution. Also important are functions like
\texttt{size} and \texttt{with\_size} that are used to control the sizes of
generated values and \texttt{fixed\_point} that is used to define recursive
generators.

The generator in Figure~\ref{fig:tree_of} uses all of these
features. It uses \texttt{fixed\_point} to define a recursive generator that
reads the current value of \texttt{size} to determine how to generate a tree.
It uses \texttt{weighted\_union} to make a random choice between an
empty tree and a node, choosing a node with weight proportional to the current
size and choosing a leaf otherwise. When generating a node, it uses
\texttt{with\_size} to reduce the value of the size parameter for future
iterations.

\subsection{Abstraction Overhead of Generator DSLs}
Just how large is the abstraction overhead of monadic generator DSLs, and where does it come from?
Figure~\ref{fig:bq-internals} shows the internals of (a simplified version of) \bq{}.
A generator \texttt{'a Bq.t} is just a function of type~\texttt{int -> SR.t -> 'a},
taking an \texttt{int} representing the current size parameter and a random seed \texttt{SR.t},
and returning a generated value \texttt{'a}.
The random seed \texttt{SR.t} from \bq's \rand{} is called \texttt{Splittable\_random}, henceforth aliased in code as \texttt{SR}.
It is an invariant of the generator library that every function of this type is
deterministic: for a fixed size and seed, it will always return the same value, so all of the randomness in testing comes from varying the initial seed.
The monad functions \texttt{return} and \texttt{bind} are defined in the usual way:
for an instance of the reader monad~\cite{moggi91}: \texttt{return} ignores the size and seed and returns its argument, while \texttt{bind} runs \texttt{g} and passes the result to \texttt{k}.
The \texttt{gen\_int} combinator simply calls out to the randomness library \texttt{Splittable\_random}, aliased as \texttt{SR} here.
Different generator DSLs use variations on this basic design, but the basics are the same across the board.

% Different PBT libraries use variations on this basic design. The largest source of variation is how the random seed is threaded through the generator.
% \bcp{Sentence starts with lowerase} \bq{} uses a reader monad with a mutable seed, while Haskell's QuickCheck uses a reader monad with an immutable state
% that is ``split'' at \texttt{bind}s. Meanwhile, ScalaCheck (a) uses an immutable state type and state monad to thread
% the state through, and (b) has an \texttt{Option} as its return type, to allow generation to fail.
% \bcp{This discussion should go elsewhere...}

\begin{figure}
\begin{lstlisting}
module Bq = struct
  type 'a t = int -> SR.t -> 'a

  let return (x : 'a) : 'a t = fun _ _ -> x

  let bind (g : 'a t) (k : 'a -> 'b t) : 'b t =
    fun size random ->
      let a = g size random in
      (k a) size random

  let gen_int (lo : int) (hi : int) : int t =
    fun _ random -> SR.int random lo hi
end
\end{lstlisting}
\caption{Internals of a Monadic Generator eDSL}
\label{fig:bq-internals}
\end{figure}

Just how much run-time overhead does this monadic abstraction introduce? To
illustrate, let's return to our running example of a constrained pair of
integers, written in both \bq{} and ScalaCheck.
Figure~\ref{fig:simplify} shows two versions of the generator, written in both languages.
The first versions (\texttt{int\_pair} in \bq{} and \texttt{intPair} in ScalaCheck) are written with the monadic generator combinators from their respective libraries.
The second versions (\texttt{int\_pair\_inlined} and \texttt{intPairInlined})
are semantically identical to the first, but have been rewritten
by (1) inlining all generator combinator definitions, and then (2)
repeatedly reducing simplifiable terms like \lstinline{(fun x -> e) e'} where an anonymous function is defined and then immediately called, to \lstinline{let x = e' in e}.

\begin{figure}[h]
  \begin{tabular}{lll}
  Library    & Generator          & Average Time per Generation (ns) \\
  \bq{}         & \texttt{int\_pair}          & 70                       \\
  \bq{}         & \texttt{int\_pair\_inlined} & 35                       \\
  ScalaCheck & \texttt{intPair}            & 458                      \\
  ScalaCheck & \texttt{intPairInlined}     & 266
  \end{tabular}
\caption{Microbenchmarks of generators in \bq{} and ScalaCheck. Average over 10,000 generations with random seeds and a fixed size.}
\label{fig:overhead-explanation-perf}
\end{figure}

The performance impact of this simplification is large (Figure~\ref{fig:overhead-explanation-perf}). In both languages, the
inlined version takes on average half as much time to generate a single pair of
\texttt{int}s.
Microbenchmarks of more realistic generators (see Section~\ref{section:eval})
show an even more dramatic performance boost.
Because the inlined versions of the generator are identical to the un-inlined
versions except for mechanical, semantics-preserving transformations,
this performance difference is attributable solely to
the different machine code generated by the compiler (and not lucky choice of random seeds).

\begin{figure}[h]
  \begin{subfigure}{.49\textwidth}
\begin{lstlisting}
let int_pair : (int * int) Bq.t =
  let%bind x = (Bq.gen_int 0 100) in
  let%bind y = (Bq.gen_int 0 x) in
  Bq.return (x,y)

let int_pair_inlined : int -> SR.t -> int * int =
  fun _ sr ->
    let x = SR.int sr ~lo:0 ~hi:100 in
    let y = SR.int sr ~lo:0 ~hi:x in
    (x,y)
\end{lstlisting}
\caption{Simplifying a generator in OCaml}\label{fig:simplify-ocaml}
  \end{subfigure}
  \begin{subfigure}{.49\textwidth}
\begin{lstlisting}
def intPair : Gen[(Long,Long)] = for {
  x <- Gen.choose(0,1000)
  y <- Gen.choose(0,x)
} yield (x,y)

def intPairInlined : (Gen.Parameters, Seed) =>
(Option[(Long,Long)],Seed) = {
 (p,seed) =>
  val (x,seed2) = chLng(0,1000)(p,seed)
  x match {
    case None => (None,seed2)
    case Some(x) =>
      val (y,seed3) = chLng(0,x)(p,seed2)
      y match {
        case None => (None,seed)
        case Some(y) => (Some(x,y),seed3)
      }
    }
  }
\end{lstlisting}
\caption{Simplifying a generator in Scala}\label{fig:simplify-scala}
  \end{subfigure}
  \caption{Simplifying Generators}\label{fig:simplify}
\end{figure}

It might be surprising to readers that this dramatic overhead exists---shouldn't the compiler perform
this simple optimization? Compilers of effectful and strict functional languages (including JIT
compilers in the case of Scala 3)
do in fact have heuristics to determine if and when to perform this particular kind of
simplification.\footnote{As we discuss in Section~\ref{section:other-langs}, purity
means that Haskell is a slightly different story. GHC can and often does
transformations of this form.} However even in cases as simple as
Figure~\ref{fig:simplify}, the indirection of \texttt{return}
and \texttt{bind} causes these heuristics to not fire.  The story is
even worse for recursive generators, as the heuristics are necessarily even more
conservative for optimizing recursive functions.

More complex
generators suffer further performance penalties due to \emph{closure
allocation}. In cases where the compiler cannot statically eliminate it,
running a monadic bind allocates a short-lived closure
for the continuation and then immediately jumps into it.
In strict functional languages like OCaml and Scala,
each individual closure allocation is relatively cheap; but doing lots of allocation in a
generator is very expensive because each allocation brings us closer to the next costly GC pause.
This effect is magnified in recursive generators: each iteration
through the recursive loop re-allocates closures for \texttt{bind}s, so the amount of allocation
per generated value scales linearly with the number of recursive generator calls.

\paragraph{Overhead of Choice Combinators}
Like \texttt{return} and \texttt{bind}, combinators like \texttt{weighted\_union} incur a performance penalty
at run time. Aside from the previously-discussed issue that compilers
cannot see through the abstraction boundary to optimize these programs,
choice combinators like \texttt{weighted\_union} come with their own particular abstraction overhead.

In practice, \texttt{weighted\_union} is (almost\footnote{
There are some generators where \texttt{weighted\_union} is passed a list which
was itself the result of a generator: the well-typed STLC term generator used in Section~\ref{section:eval} is an example of this.
}) always called with an explicitly constructed list, as in
\texttt{weighted\_union [(w1,g1); (w2,g2); (w3,g3)]}.
This is because the most common use case for \texttt{weighted\_union} is
to choose between one of the different constructors of an algebraic datatype,
the options for which are always known.
This list is allocated at the call site and then never needed after the call to \texttt{weighted\_union} returns.
Since the elements of the list and its length \texttt{n} are known,
a compiler could in principle unroll the loops in \texttt{weighted\_union} to depth \texttt{n} and specialize the function at each call site to avoid allocating the list.
Unfortunately, (almost\footnote{GHC is a notable exception here, performing sophisticated list fusion optimizations \cite{gill93}.}) no compilers perform this kind of optimization.
This allocation (or rather, its tendency to cause GC pauses)---as well as the cost of running the code
to traverse arbitrary lists compared to unrolled loops---has a significant impact on
performance.

Last, many PBT libraries---including both \bq and ScalaCheck---implement their \texttt{weighted\_union} combinators
in ways that are asymptotically more efficient but slower in common cases than a more naive algorithm.
Weighted union uses the Fitness Proportionate Selection algorithm~\cite{geneticalgorithm}, which (1) samples a number $r$
between $0$ and the sum of the weights, and then (2) finds the first generator for which the cumulative
weights in the list before it exceeds $r$. This second part can be accomplished in $O(\log_2 n)$ time by
a binary search. However, since the lists are short in practice, a linear scan is almost always faster.
Moreover, both \bq{} and ScalaCheck allocate auxiliary data structures (an array in \bq{} and a BST in ScalaCheck) to
perform this search, which incurs further run-time overhead.

% \subsubsection{Function call overhead}
\subsection{Inefficient Randomness Libraries}
\label{subsection:ineff-rand}
The core of any PBT generator library is a source of randomness.
Different PBT libraries use different randomness libraries implementing different algorithms.\footnote{The common term for such an algorithm or library is a ``Random Number
Generator'' (RNG). We will avoid this term and instead say ``randomness library''
to avoid confusing RNG implementations with the PBT generator libraries that
use them.}
Following the original Haskell QuickCheck implementation, \bq\ uses the SplitMix
algorithm~\cite{splitmix}, implemented by \texttt{Splittable\_random}. Meanwhile,
ScalaCheck uses the JSF algorithm~\cite{jsf}.
% https://www.pcg-random.org/posts/bob-jenkins-small-prng-passes-practrand.html
% https://burtleburtle.net/bob/rand/smallprng.html

The randomness library is the hottest part of the hot path.
Indeed, even basic generators---like ones generating a single \texttt{int} or \texttt{float} uniformly within a range---can sample \emph{unboundedly many}
random numbers, since they usually use versions of rejection sampling~\cite{vn51} to find a value within the range.
Moreover, generator combinators like \texttt{list} usually make $O(n)$ calls to the \texttt{int} or \texttt{float} generators.
Because of this, the speed of a single sample matters a great deal. Unfortunately,
while existing PBT libraries by and large make sensible choices for their randomness libraries,
they are not chosen with performance in mind, leading to worse bug-finding power than what is possible.

For example, significantly faster algorithms than SplitMix or JSF exist,
such as the Lehmer algorithm~\cite{lehmerrng}, WyRand~\cite{wyrand}, and the xorshiro family of algorithms~\cite{xorshiro}.
These all run between 1.2-1.6x faster per byte than SplitMix in microbenchmarks~\cite{lemiremicrobenchmarks}.
Plenty of other known optimizations on top of algorithm choice could also be implemented,
including pipelined or ahead-of-time sampling.
% https://github.com/lemire/testingRNG/tree/master
% https://github.com/wangyi-fudan/wyhash?tab=readme-ov-file

Of course, simply arguing that the randomness library is on the hot
path for PBT does not guarantee that a faster sampling leads to measurably faster
generation; for that, we need an experiment. The most obvious experiment is to
simply swap out the randomness library of either \bq{} or ScalaCheck one of the aforementioned faster algorithms.
But, as discussed previously, \camlname{} and \scalaname{} are designed to be 100\% semantically equivalent replacements for
their unstaged counterparts, ensuring that any bug-finding speedups are solely attributable to generator performance and not lucky seeds.
Choosing a different \rand{} would break this property, making it much more challenging to assess the performance compared to the baseline.
To get around this, we exploit a coincidence:
\texttt{Splittable\_random}---the OCaml implementation of SplitMix that \bq{} uses---is slow in a way that
can be improved \emph{without} changing the algorithm.
In particular,
due to implementation details related to the OCaml garbage collector,
values of the OCaml type \texttt{int64} are not machine words, but rather
\emph{pointers} to machine words. This means
that \emph{all} \texttt{int64} operations (both arithmetic and bitwise) must
allocate memory cells to contain their output, which has a significant performance penalty.
By building a version that uses much faster ``unboxed'' 64-bit integer
arithmetic, we (a) demonstrate how much faster bugs can be found just by
using a more performant \rand{}, and (b) explore which generators benefit the most from faster \rands{}.

\section{How Can I Make My Generator Library Faster?}
\label{section:faster-generators}

With a sense of two main inefficiences in generator libraries, we set out in this section to investigate solutions.
We begin with a quick tour through multi-stage programming (Section~\ref{subsection:msp}),
then incrementally build \camlname, layering on features and
functionality  (Sections~\ref{subsection:basic-design} through \ref{subsection:type-derived}).
We begin with simple staging tricks, then move on to more advanced techniques that
leverage our domain-specific understanding of PBT generators.
Last, we present a controlled experiment demonstrating how a
faster \rand{} leads to failing faster  (\ref{subsection:faster-rng}).

\subsection{Background: Multi-Stage Programming}
\label{subsection:msp}

In multi-stage programming, or simply ``staging,'' programs execute
in multiple stages, with each stage producing code to be run
in the next. For the purposes of this paper, we need just two stages: compile time and run time.

\paragraph{Staging eDSLs}
One of the primary uses of staging is in embedded DSL construction \cite{sheard99, lms, trattdsl}.
While embedded DSLs are powerful tools, they have a well-known drawback:
the functional abstractions used to build eDSLs tend to prevent compilers from generating efficient machine code.
This is \emph{abstraction overhead} \cite{carrette05, moller20}: the layers of abstraction that
make eDSLs pleasant to use are precisely what prevents them from being fast.
Indeed, the root causes of many performance issues we discovered in Section~\ref{section:motiv} are
not unique to generator DSLs, but pervasive across eDSLs.
In light of this issue, staging is often used as a lightweight compiler for DSLs. The compile-time evaluation stage
transforms the DSL code, eliminating abstractions to produce code that the host language compiler can generate fast machine code for.
This recipe has been used to great effect to stage eDSLs for stream processing \cite{moller20,strymonas}, parser combinators \cite{sspc, staged-parsers, krishnaswami19, flap},
and query processing \cite{rhyme}.

Many languages have some degree of staging functionality, including
Scala 3~\cite{scalamacros}, Haskell~\cite{templatehaskell}, Racket~\cite{flatt12}, OCaml~\cite{metaocaml,macocaml}, and Java~\cite{mint}.
For this paper, we have implemented staged PBT libraries in both OCaml (using MetaOCaml~\cite{metaocaml}) and Scala 3 (using
\texttt{scala.quoted}). All staged code presented in the body of this paper is \camlname{} code written in OCaml;
\scalaname{} is very similar.
We discuss the potential for staged generator DSLs in other languages in Section~\ref{section:other-langs}.

\paragraph{Staging in MetaOCaml}
MetaOCaml's staging functionality is exposed through the type \texttt{'a code}. A
value of type \texttt{t code} is an OCaml term
of type \texttt{t}.
Values of \texttt{code} type are introduced by \emph{quotes} (written
\texttt{.<$\ldots$>.}). Quotes delay execution of a program until run time. For
example, the program \texttt{.< 5 + 1 >.} has type \texttt{int code}.  Note that
this is not the same as \texttt{.< 6 >.}: because brackets delay computation,
the code is not \emph{executed} until the next stage (run time).
Values of type \texttt{code} can be combined using an \emph{escape},
written \texttt{.\~{}(e)} (or just \texttt{.\~{}x}, when \texttt{x} is a variable).  Escaping lets you take a value of type
\texttt{code} and ``splice'' it directly into a quote.  For example, the
program \texttt{let x = .<1 * 5>. in .< .\~{}(x) + .\~{}(x) >.} evaluates to
\texttt{.<(1 * 5) + (1 * 5)>.}  MetaOCaml enforces correct scoping and macro
hygiene, ensuring that variables are not shadowed when open terms are
spliced in under binders.

The power of staging for optimizing away abstraction overheads comes from
defining functions that accept and return \texttt{code} values.  A function
\texttt{f : 'a code -> 'b code} takes a program computing a
run-time \texttt{'a} and transforms it into a program computing a run-time
\texttt{'b}. In particular, because \texttt{f} itself runs at compile time, the
fact that the programmer called \texttt{f} does not matter at run time---the abstraction
that \texttt{f} defines has been eliminated.
A code-transforming function \texttt{'a code -> 'b code} can also be converted to \emph{code for a function}---a value of type
\texttt{('a -> 'b) code}---with the following program:
\begin{lstlisting}
let eta (f : 'a code -> 'b code) : ('a -> 'b) code = .<fun x -> .~(f .<x>.)>.
\end{lstlisting}

This program is known as ``The Trick'' in the partial evaluation and multi-stage programming literature~\cite{metaocamlimpl}.
% USE THIS: https://arxiv.org/pdf/2309.08207
It returns code for a function that takes an argument \texttt{x} and splices in the result of calling \texttt{f} on just
the quoted \texttt{x}.
For example, the following program
\begin{lstlisting}
let is_even x = .< .~x mod 2 == 0 >. in
let succ x = .< 1 + .~x >. in
eta (fun x -> succ (is_even x))
\end{lstlisting}
reduces at compile time to \texttt{.< fun x -> (1 + x) mod 2 == 0 >.}
By composing the two code-transforming functions together at compile time and only then turning them into a run-time function,
the two functions are fused together.
This is the basis of how staging is used to eliminate the abstraction overhead of DSLs.
By writing DSL combinators as compile-time functions---and calling \texttt{eta} at the end on the completed DSL program---we
can ensure that any overhead of using the combinators is eliminated before run time.

% \jwc{
%   \begin{itemize}
%     \item The \texttt{'a code} type, quote, escape or ``splice'', stage distinction.
%     \item MetaOCaml ensures correct scoping and hygene by preventing alpha-collision when you splice open terms.
%     \item The difference between \texttt{('a -> 'b) code} and \texttt{'a code -> 'b code}
%     \item We can convert one way, but not the other.
%     \item Functions \texttt{'a code -> 'b code} ``fuse''.
%     \begin{itemize}
%       \item Consider writing \texttt{even . succ}. This
%       \item Consider \texttt{even\_c : int code -> bool code = fun cx -> .<.~cx mod 2 == 0>.} and \texttt{succ\_c : int code -> int code = fun cx -> .< .~x + 1 >.}
%       \item If you do \texttt{to\_dyn (even\_c . succ\_c)}, you get \texttt{fun x -> (x + 1) mod 2 == 0}. Composing functions from code to code, and then \emph{only at the end} stamping out
%       a dynamic function value eliminates the function abstraction.
%     \end{itemize}
%     \item This is the basis of (WORD). By defining a library with functions with types like \texttt{'a code -> 'b code}, we can ensure that the
%     abstractions the library introduces are fully eliminated at compile time.
%   \end{itemize}
% }

\subsection{Design of a Staged Generator DSL}
\label{subsection:basic-design}
To build a staged version of a generator DSL, we want to rewrite the generator
combinators to do as much as possible at compile time.
The compile-time stage will then produce simplified code, free of any DSL abstraction, that can be compiled and run
with different sizes and seeds.
Our job is thus to carefully bisect the DSL, determining which inputs to generator
combinators are known statically (and can be part of the compile time stage) and
which parts are only known at run time (and must be treated as \texttt{code}).
In the staging literature, this task is known as ``binding-time
analysis\cite{jones1993partial}.''

% carefully break up so that you push what you can into the compile time stage.
% Figuring out which parts of the generator are known statically ()
% Recall \jwc{did we talk about this} that staging a DSL involves changing the
% combinators to run at compile time by carefully annotating their types with \texttt{code}s.
% Deciding which types \texttt{t} can
% instead be \texttt{t code}---in other words,
% figuring out which parts of the DSL can be determined statically (and can be
% part of the compile-time stage), and which parts are only known dynamically (and
% hence must be \texttt{code})---is an art known as ``binding-time analysis''

The crux of our binding time analysis is that the \emph{only} parts of a
generator that are not
known at compile time are the random seed and
size parameters. In practice, generators themselves are entirely known at compile time.
This leads us to define our library's generator type \texttt{'a
  Gen.t} as
\[
\texttt{'a Gen.t} \ \ = \ \  \texttt{int code -> SR.t code -> 'a code}.
\]
That is, a \texttt{Gen.t} is a compile-time
function from dynamically known size and seed to dynamically determined result. 

The basic functionality of the staged monadic generator DSL can be
found in Figure~\ref{fig:gen-staged-basic}.

\begin{figure}[h]
  \begin{lstlisting}
  module Gen = struct
    type 'a t = int code -> Random.t code -> 'a code

    let return (cx : 'a code) : 'a t = fun size random -> cx

    let bind (g : 'a t) (k : 'a code -> 'b t) : 'b t =
      fun size random ->
        .<
          let a = .~(g size random) in
          .~(k .<a>. size random)
        >.

    let int (lo : int code) (hi : int code) : int t =
      fun size random ->
        .< SR.int .~random .~lo .~hi >.

    let to_bq (g : 'a code Gen.t) : ('a Bq.t) code =
      .<
        fun size random -> .~(g .<size>. .<random>.)
      >.
  end
  \end{lstlisting}
  \caption{Basic Staged Generator Library\ifaftersubmission\bcp{We could
      save some space if needed by eliminating a couple of linebreaks
      here.}\fi}
  \label{fig:gen-staged-basic}
  \end{figure}

The constant generator \texttt{return} runs at compile time. Given \texttt{cx : 'a code}, the code for an \texttt{'a}, it
returns the generator that ignores its \texttt{size} and
\texttt{random} arguments and simply returns \texttt{cx}. Similarly,
\texttt{bind g k} sequences generators by passing the result of running the generator \texttt{g} to the continuation \texttt{k}. However, instead of getting access
to the particular value generated by \texttt{g}, the continuation
\texttt{k} only gets access to \texttt{code} for the value sampled from \texttt{g}:
we know that at run time the code \texttt{g} generates will produce \emph{some} \texttt{'a}, but at compile time we cannot inspect the value.
Operationally, bind takes code for the size and seed and returns code that (1) let-binds a variable \texttt{a} to spliced-in code of \texttt{g} and then
(2) runs the spliced-in continuation \texttt{k}.
Both function applications \texttt{g size random}
and \texttt{k .<a>. size random} run at compile time.
\texttt{Gen.int} is the generator that samples an int from the randomness library.
Given any size and random seed, it returns a code block that calls \texttt{SR.int} with that random seed. Because the lower and upper bounds
might not be known at compile time---they may themselves be the results of calling \texttt{Gen.int}---the arguments
\texttt{lo} and \texttt{hi} are of type \texttt{int code} and get spliced into the code block as arguments to \texttt{SR.int}.
Lastly, \texttt{to\_bq} turns a staged generator into code for a normal \bq\
generator. This is just a two-argument version of ``The Trick''
from Section~\ref{subsection:msp}.
\ifaftersubmission
\hg{TODO: Be really careful with formatting of the above paragraph. It's very
easy for some inline code to get split weirdly over a line break or just be
difficult to parse in general, and that could throw the reader off when the
content is already pretty low-level}
\fi

Returning to our running example, Figure~\ref{fig:running-staged} shows the
int-pair generator written with the staged \texttt{Gen.t} monad, as well as the
inlined code that results from calling \texttt{Gen.to\_bq} (changing some identifier names for clarity).
The code generated is identical to the manually inlined version from Section~\ref{section:motiv}.

\begin{figure}[h]
\begin{lstlisting}
let int_pair_staged : (int * int) Gen.t =
  Gen.bind (Gen.int .<0>. .<100>.) (fun cx ->
    Gen.bind (Gen.int .<0> cx) (fun cy ->
      Gen.return .<(.~cx,.~cy)>.
    )
  )

let int_pair : (int * int) Bq.t code = Gen.to_bq int_pair_staged
(* int_pair = .< fun size random ->
                 let x = SR.int random 0 100 in
                 let y = SR.int random 0 x in
                 (x,y)
               >.   *)
\end{lstlisting}
\caption{Pairs of Ints, Staged}
\label{fig:running-staged}
\end{figure}

\subsection{PBT-Specific Staging}
While the techniques in the previous section eliminates monadic
DSL overhead using standard ideas in staging, it cannot address some of the
more subtle and domain-specific performance issues identified in Section~\ref{section:motiv},
like hot-path allocations induced by generator combinators such as \texttt{weighted\_union}.
Removing these allocations requires an observation that transcends simple 
inlining: though the \emph{elements} of the lists may be dynamic, their
\emph{spine}---the structure of the list itself---is in practice always statically
known. This is an instance of \emph{partially-static data} \cite{partially-static}:
a data structure whose shape is static but whose contents are dynamic.

This observation enables an optimization unique to the PBT generator library setting.
Because the spine is static, we can unfold the choice-selection
logic over the list at compile time time, specializing \texttt{weighted\_union} to the exact number
and weights of choices. The result is a specialized generator that branches directly to the chosen
alternative, without ever materializing the list of choices at run time.

% In Section~\ref{section:motiv}, we noted that generator combinators like
% \texttt{weighted\_union} allocate lists in the hot path of the generator. Even
% though these lists are usually small---at most a few dozen elements---each
% allocation takes us closer to the next garbage collection, which is bad for
% performance.

% The simple staging we've used thusfar to remove the overhead of \texttt{return}
% and \texttt{bind} cannot eliminate this allocation --- we must use structure specific
% to PBT generator combinators to our advantage.

% In particular, the \emph{spine} of the list of choices that \texttt{weighted\_union} picks from is
% in practice always statically known, even if the elements themselves may be determined dynamically
% --- the list is \emph{partially-static data} \cite{partially-static}.
% Because of this, we can statically unroll the outer loop of the \texttt{weighted\_union} combinator
% and specialize to the potential choices at compile-time.

% This is an ideal opportunity to exercise another feature of staging: compile-time specialization.

% Since we almost always know the particular list of choices at compile time, a staged version of \texttt{weighted\_union}
% can generate \emph{different code} depending on the number of generators in the union.
% If we use weighted union on a compile-time list of generators \texttt{g1}, \texttt{g2}, and \texttt{g3},
% we can emit code that picks between the generators without realizing the list at
% run time.

\begin{figure}[h]
\begin{lstlisting}
module Gen =
$\dots$
  let pick (acc : int code) (weighted_gens : (int code * 'a t) list) size random : 'a code =
    match weighted_gens with
    | [] -> .< failwith "Error" >.
    | (wc,g) :: gens' ->
      .<
        if .~acc <= .~wc then .~(g size random)
        else
          let acc' = .~acc - .~wc in
          .~(pick .<acc'>. gens' size random)
      >.

  let weighted_union (weighted_gens : (int code * 'a t) list) : 'a t =
    let sum_code = List.foldr (fun acc (w,_) -> .<.~acc + .~w>. ) .<0>. weighted_gens in
    fun size random ->
      .<
        let sum = .~sum_code in
        let r = SR.int .~random_c 0 sum in
        .~(pick .<r>. weighted_gens size random)
      >.
\end{lstlisting}
\caption{Staged Weighted Union}
\label{fig:staged-weighted-union}
\end{figure}

Figure~\ref{fig:staged-weighted-union} shows the code for such a staged weighted union.
Crucially, it takes a \emph{compile-time} list \texttt{weighted\_gens} of
generators and weights. The weights themselves might only be known at run time---it is
common to use the current size parameter as a weight, for instance---so they are \texttt{code}s.
Instead of building a data structure representing a histogram of the distribution described by the weights at run time and then traversing it,
the compile-time \texttt{weighted\_union} combinator generates a
tree of \texttt{if}s that picks out the selected generator.

% \texttt{Gen.weighted\_union} begins by computing \texttt{sum\_code}, an \texttt{int code}
% that is the sum of the weights. Note that this happens at compile time: we fold over a list
% known at compile time to produce another code value.
% We then call \texttt{SR.int} to sample a random number \texttt{r} between \texttt{0} and the sum.
% Finally, we splice in the result of calling the helper function \texttt{pick}. \texttt{pick}
% produces a tree of \texttt{if}s by again traversing the list of generators at compile time.
% This tree of \texttt{ifs} ``searches'' for the generator corresponding to the
% sampled value \texttt{r}, and then runs it.
% \hg{I'm not sure this helps me. I think I'd prefer a less detailed explanation
% of the code that conveys the intuition and let the reader actually read the code
% if they want to know specifics. As it stands the explanation is just too dense
% for me to process}

Figure~\ref{fig:staged-weighted-union-example} demonstrates a use of this staged weighted union.
Given a list of (in this case constant) generators with weights ``the current size parameter,'' \texttt{2}, and \texttt{1},
the generated code first computes the sum of these numbers, samples between \texttt{0} and the sum, and
traverses the tree of three \texttt{if}s to find the correct value to return.

\begin{figure}[h]
\begin{lstlisting}
let grades : char Bq.t = Gen.to_bq (
  Gen.bind size (fun n ->
    Gen.weighted_union [
      (n, Gen.return .<'a'>.);
      (.<2>., Gen.return .<'b'>.);
      (.<1>., Gen.return .<'c'>.);
    ]
  )
)
(*
.< fun size random ->
    let sum = size + 2 + 1 + 0 in
    let r = SR.int random 0 sum in
    if r <= size then 'a' else
      let r' = r - size in
      if r' <= 2 then 'b' else
        let r'' = r' - 2 in
        if r'' <= 1 then 'c' else failwith "Error"
>.
*)
\end{lstlisting}
\caption{Use of Staged Weighted Union}
\label{fig:staged-weighted-union-example}
\end{figure}

\subsection{Let-Insertion and Effect Ordering}
Careful readers might note that the definition of \texttt{bind} (Figure~\ref{fig:bind-with-let-binding})
was more complicated than one might expect. Why not define bind in the standard way
for a reader monad (Figure~\ref{fig:bind-without-let-binding})?
Unfortunately, the latter definition is wrong in our context as it leads to incorrect code being generated.
For example, consider \texttt{Gen.bind' (Gen.int .<0>. .<1>.) (fun x -> Gen.return .<(.~x,.~x)>.)}.
This generates the run-time code \texttt{fun size random -> (SR.int random 0 1, int SR.random 0 1)},
which is incorrect.  Instead of generating a single integer and returning it
twice, it samples two different integers.
This matters because, as described in Section~\ref{section:intro}, \camlname and
\scalaname are intended to be equivalent to their unstaged counterparts.

\begin{figure}[h]
  \begin{subfigure}{.45\textwidth}
\begin{lstlisting}[xleftmargin=0pt, xrightmargin=-3pt, aboveskip=0pt, belowskip=0pt]
let bind (g : 'a t) (k : 'a code -> 'b t) : 'b t =
    fun size random ->
      .<
        let a = .~(g size random) in
        .~(k .<a>. size random)
      >.
\end{lstlisting}
\caption{\texttt{bind}, with a let-binding}\label{fig:bind-with-let-binding}
  \end{subfigure}
  \begin{subfigure}{.45\textwidth}
\begin{lstlisting}[xleftmargin=0pt, xrightmargin=-27pt, aboveskip=0pt, belowskip=0pt]
let bind' (g : 'a t) (k : 'a code -> 'b t) : 'b t =
    fun size random -> k (g size random) size random
\end{lstlisting}
\caption{\texttt{bind'}, the ``standard'' bind for the reader monad}\label{fig:bind-without-let-binding}
  \end{subfigure}
  \caption{\texttt{bind}, two ways}
  \label{fig:bind-two-ways}
\end{figure}

In essence, the behavior of splice \texttt{.\~{}cx} in a staged function \texttt{f(cx : 'a code) = ...} is to \emph{copy}
the entire block of code, effects and all. To ensure that the randomness effects of the first generator are executed only once
but that the value can be used in the continuation multiple times, the correct \texttt{bind} let-binds the result of generation to a variable
and then passes it to the continuation.

% The library \jwc{ensure this is consistent with the way we talk about the project, cf cross-language} is careully designed to
% preserve exactly the effect order of base quickcheck \jwc{and the scala version to preserve the effect ordering of ScalaCheck}.
% \jwc{This fact should go earlier. Not really sure about where this subsection should actually land, but it needs to be said somehwere.}
% \hg{+1, I think this feels sort of out of place here, but I also think moving it
% up would make that part harder to understand too. Is there a way to tie this in
% to our discussion about maintaining the precise set of choices? The two issues
% are different, but they're related}

\subsection{CodeCPS and a Monad Instance}
\label{subsection:codecps}
Another subtle issue prevents the version of the library design discussed so far
from being used as a drop-in replacement for an existing generator DSL: the types of \texttt{return : 'a code -> 'a Gen.t}
and \texttt{bind : 'a Gen.t -> ('a code -> 'b Gen.t) -> 'b Gen.t} aren't quite right.
For the type \texttt{'a Gen.t} to actually be a monad,
these types cannot mention \texttt{code}. This is not just a theoretical issue; it is a significant usability concern.
The syntactic sugar for monadic programming (\texttt{let\%bind} in OCaml, \texttt{foreach} in Scala, \texttt{do} in Haskell, etc)
that makes it smooth can \emph{only} be used if
return has type \texttt{'a -> 'a Gen.t} and bind has type \texttt{'a Gen.t -> ('a -> 'b Gen.t) -> 'b Gen.t}

To support convenient monadic programming, we need to adjust the type of \texttt{'a Gen.t} slightly.
An initial attempt is to try \texttt{type 'a t = int code -> SR.t code -> 'a}. If we strip the \texttt{code} off the result type,
the functions \texttt{return (x : 'a) = fun \_ \_ -> x} and \texttt{bind g k = fun size seed -> k (g size seed) size seed}
have the proper types for a monad instance. Then, any combinators of type \texttt{'a Gen.t} become \texttt{'a code Gen.t} in this new version.

However, this definition of bind doesn't have call-by value effect semantics, as
discussed in the previous section.  And because the type of \texttt{g size seed} is
now just \texttt{'a} (not necessarily \texttt{'a code}), we cannot perform the
let-insertion needed to preserve the CBV effects. To solve this problem, we
turn to a classic technique from the multistage programming literature: writing
our staged programs in continuation-passing style \cite{bondorf92}.

\begin{figure}[h]
  \begin{lstlisting}
  module CodeCps = struct
    type 'a t = { cps : 'z. ('a -> 'z code) -> 'z code }

    let return x = {cps = fun k -> k x}

    let bind (x : 'a t) (f : 'a -> 'b t) : 'b t =
      {cps = fun k -> x.cps (fun a -> (f a).cps k)}

    let run (t : ('a code) t) : 'a code = t.cps (fun x -> x)

    let let_insert (cx : 'a code) : 'a code t =
      {cps = fun k -> .< let x = .~cx in .~(k .<x>.) >.}
  end

  module Gen = struct
    type 'a t = int code -> SR.t code -> 'a CodeCps.t

    let return (x : 'a) : 'a t = fun _ _ -> Codecps.return x

    let bind (g : 'a t) (f : 'a -> 'b t) =
      fun size random ->
        CodeCps.bind (g size random) (fun x -> (f x) size random)

    let int (lo : int code) (hi : int code) : int code t =
      fun size random -> let_insert .< SR.int .~random .~lo .~hi >.
  end
  \end{lstlisting}
  \caption{CodeCPS and the Final Gen Monad}
  \label{fig:codecps-and-final-gen}
  \end{figure}

In Figure~\ref{fig:codecps-and-final-gen}, following prior work \cite{kovacs24, carrette05}, we define
the type \texttt{'a CodeCps.t = 'z. ('a -> 'z code) -> 'z code}: a polymorphic continuation transformer with the result type
always in \texttt{code}.\footnote{This is an instance of the \emph{codensity} monad \cite{janis08}---a fact that deserves further investigation.}
The monad instance for this type is the standard instance for a CPS monad with polymorphic return type.
In prior work, this type is often referred to as the ``code generation'' monad.
This is because a value of type \texttt{('a code) CodeCps.t} is like an ``action'' that generates \texttt{code}:
\texttt{CodeCps.run} passes the continuation transformer the identity continuation to produce a value of type \texttt{'a code}.
To avoid confusion with random data generators, we refer to this type as \texttt{CodeCps.t}.
Most importantly, the \texttt{CodeCps} type supports a function \texttt{let\_insert}, which, given \texttt{cx : 'a code},
let-binds \texttt{let x = .\~{}cx}, and then passes \texttt{.<x>.} to the
continuation.
% \hg{I'm getting lost in the inline code again}

We can then redefine
our staged generator monad type to be
\texttt{'a Gen.t = int code -> SR.t code -> 'a CodeCPS.t}, as shown in Figure~\ref{fig:codecps-and-final-gen}.
The (old) type \texttt{'a Gen.t} is now written as the (new) type \texttt{'a code Gen.t}, and this type change
carries through all of our combinators. For example, \texttt{Gen.int} now returns \texttt{int code Gen.t}.

This approach gives us the best of both worlds. First, we get a monad instance for \texttt{'a Gen.t} with the correct types, which lets us
use the monadic syntactic sugar of our chosen language. Moreover, we also get to maintain the correct
effect ordering: effectful combinators like \texttt{Gen.int} do their \emph{own} let-insertion, ensuring
that a program like \texttt{Gen.bind Gen.int (fun x -> ...)} generates a let-binding for the result of sampling the randomness library.
For example, \texttt{bind (int .<0>. .<1>.) (fun cx -> return .<(.~cx,.~cx)>.)} now correctly generates
\texttt{.< fun size random -> let x = SR.int random 0 1 in (x,x) >.}
This design is less obviously correct, and does require some care. Rather than \texttt{bind} ensuring correct evaluation order once and for all,
individual combinators must be carefully written to ensure that \texttt{'a code} values that contain effects are \texttt{let\_insert}ed.
To validate the library, we built a PBT harness to compare staged generators to their \bq equivalents over 1,000 random seeds.
By differentially testing~\cite{difftesting} a large suite of generators in this way, we gained confidence that \camlname is equivalent to \bq{}.

% \subsubsection{The Trick at Other Types}
% To write more interesting generators, we also need the ability to generate code that manipulates run-time values.
% For instance, consider this generator

% \jwc{need a better example here}

% \begin{figure}
% \begin{lstlisting}
% let int_or_zero : int Bq.t =
%   let%bind n = size in
%   if n <= 5 then return 0 else Bq.int

% let split_bool (b : bool code) : bool Gen.t =
%   fun _ _ ->  _

% \end{lstlisting}
% \caption{??}
% \label{fig:trick-example}
% \end{figure}

% \jwc{Describe split---this section is approximately experts-only, but you can just point at the Andras paper and the things it cites.}

\subsection{Recursive Generators}
Generating values of recursive datatypes requires recursive generators. Different generator DSLs support recursive generators differently. Some allow recursive generators
to be defined as recursive functions, while others
(including both \bq and ScalaCheck) expose a fixed-point combinator to construct recursive generators.
Given a step function that takes a ``handle'' to sample from a recursive generator call, it
ties the knot and builds a recursive generator.

In our setting, letting programmers define recursive generators as recursive functions is out of the question.
With staged programming, recursion must be handled with care: it is far too easy to accidentally recursively
define an infinite \texttt{code} value and have the program diverge at compile time, when trying to write
a \texttt{code} representing a recursive program.
To this end, we develop a staged recursive generator combinator\footnote{
We actually provide a more general API that allows programmers to define \emph{parameterized}
recursive combinators, of type \texttt{'r code -> 'a code Gen.t}, for any type \texttt{'r}.
}, whose API is shown in Figure~\ref{fig:staged-recursive-generator}.
The recursion API consists of an opaque type \texttt{'a handle}, and a function \texttt{recurse} to perform recursive calls.
Programmers can then define recursive generators by \texttt{fixed\_point}, which ties the recursive knot.

% \begin{lstlisting}
%   type ('a,'r) handle = 'r code -> 'a code t

%   let recurse (f : ('a,'r) handle) (x : 'r code) : 'a code t =
%     fun size random ->
%       Codecps.bind ((f x) size random) @@ fun c ->
%       Codecps.let_insert c

%   let recursive (step : ('a,'r) handle -> 'r code -> 'a code t) (x0 : 'r code) : 'a code t =
%     fun size_c random_c ->
%       Codecps.return @@
%         .< let rec go x size random = .~(
%               Codecps.code_generate @@
%               (step (fun xc' -> fun ~size' ~random' -> Codecps.return .< go .~xc' .~size' .~random' >.) .<x>.) .<size>. .<random>.
%             )
%           in go .~x0 .~size_c .~random_c
%         >.
% end
% \end{lstlisting}

\begin{figure}[h]
\begin{lstlisting}
type 'a handle
val recurse : 'a handle -> 'a code Gen.t
val fixed_point : ('a handle -> 'a code Gen.t) -> 'a code Gen.t
\end{lstlisting}
\caption{Staged Recursive Generator Combinator API}
\label{fig:staged-recursive-generator}
\end{figure}

\subsection{Staging Type-Derived Generators}
\label{subsection:type-derived}

Generators are traditionally handwritten, but some PBT libraries allow users to synthesize
them automatically from type definitions. Type-derived generators are
convenient---the derivation process requires no manual effort---but also limited:
they are unable to account for constraints not encoded in the type. For example, they can
generate arbitrary trees, but not binary search trees. When such constraints are present,
type-derived generators are usually less effective than hand-crafted ones, since most
generated values will be invalid.

The speed of generation becomes particularly important in this setting. In a generator
that produces only valid values, only a subset of them will trigger a bug; in
a type-derived generator, however, only a subset of generated values will be valid,
and only a subset of \textit{those} will find a bug. As a result, many more values must be produced,
making generation speed particularly crucial.

The type-deriving algorithm follows a compositional pattern. Generators for complex types
are synthesized by structurally composing generators for their subtypes:
base types are mapped to primitive generators included in the generator library;
product types such as tuples and records are handled by sampling each component
using \texttt{bind} and aggregating the results; sum types, or variants, are generated using
a \texttt{weighted\_union} of the generators for each case; for recursive types,
the entire generator is wrapped in a fixed-point combinator and a recursive handle
is used as the generator for all recursive occurrences in the type.
This compositional approach maps naturally to staged generation:
to derive a staged generator, we replace each standard combinator
with its staged counterpart. The derivation algorithm remains unchanged.

Our implementation in OCaml uses the PPX (PreProcessor eXtension) system to synthesize
staged generators from type definitions. However, any language that supports type-derived
generators can implement a staged version using a similar implementation to the original. In Scala,
for instance, the same strategy could be realized using type class
resolution. In both settings, the result is a three-stage process: a
metaprogram---via PPX or type classes---constructs a generator expression
composed of staged combinators; this expression is evaluated at compile
time to produce a specialized generator; and finally, the generator is
executed at run time to produce values.

We implemented this in \camlname; it is not yet implemented in \scalaname. We benchmark the results in \secref{section:eval}.

\subsection{Performance Opportunities in Randomness Libraries}
\label{subsection:faster-rng}
% \jwc{Unclear what we should call this section}
% \bcp{``Performance opportunities in the RNG?''}
As we discussed in Section~\ref{subsection:ineff-rand}, choosing an inefficient randomness library is
another impediment to finding bugs fast.
To demonstrate that faster random sampling can significantly impact bug-finding power, we
use the controlled experiment suggested by OCaml's inefficient implementation of
SplitMix. By replacing this relatively slow randomness library with a faster but semantically equivalent
implementation, we can precisely quantify the bug-finding speedup that a better randomness library gives across a range of PBT scenarios.

Note that the performance intervention described here is OCaml-specific.
In most other PBT frameworks, the randomness library used operates on machine
integers, so this \emph{particular} inefficiency does not exist.
However, the insights that we will derive from this experiment in Section~\ref{section:eval}---
about which kinds of generators benefit the most from faster randomness and how it leads to faster bug-finding---
are applicable to all languages.

The precise details of the SplitMix algorithm~\cite{splitmix} are not important for
present purposes; the key fact is that all of its operations are
defined in terms of arithmetic and bitwise operations on 64-bit integers. In OCaml,
because of details related to the garbage collector, the 64-bit integer type \texttt{int64} is represented
at run time as a \emph{pointer} to an unscanned block of memory containing (among other things)
a 64-bit integer~\cite{realworldocaml}. This means that all operations that return an \texttt{int64} must allocate this block of memory,
which has a significant impact on performance. A single call to
one of the \bq library functions---like generating an arbitrary integer---may call into the \texttt{Splittable\_random}
library many times. Each call into \texttt{Splittable\_random}
may sample many times from the core SplitMix sampling routine \texttt{next\_int64}, i.e., using rejection sampling to find a value within a range.
Finally, each call to \texttt{next\_int64} allocates 9 words, and each allocation brings us closer
to the next garbage collection pause. While small allocations like these are \emph{very} fast to perform and subsequently
collect in OCaml\footnote{The OCaml GC is a generational collector~\cite{realworldocaml}, and since these allocations are small and mostly very short lived, they will all be minor allocations, never to be promoted.}
, we will see in Section~\ref{section:eval} that they can still have
a large performance impact on generators that spend most of their time sampling data.
To avoid these allocations, we
reimplement SplitMix in C and call out to it with the OCaml FFI. The C version of the library uses
proper \texttt{int64\_t} arithmetic, only boxing and unboxing integers at the call boundaries between OCaml and
C code.\footnote{Ideally in the future, one would not need to call out to C for this: the Jane Street bleeding-edge OCaml compiler has support
for unboxed types~\cite{unboxedtypes}, which (among other things) would let us implement a version of SplitMix that does not allocate, directly
in OCaml. Unfortunately, the Jane Street branch of the compiler is incompatible with MetaOCaml, which we use to implement the metaprogramming
discussed in the previous sections.}

\section{Evaluation}
\label{section:eval}
% \jwc{Initial section here about eval}\tr{todo}
We evaluate the raw generator performance and bug-finding speed of
\name generators across a range of benchmarks. Our experiments compare generators
built using our technique---with and without our improved SplitMix (``\csplitmix'')---against those implemented with
existing generator libraries and their default randomness mechanisms. Specifically,
we implement semantically
equivalent staged generator libraries that replicate the behavior of \bq
in OCaml and ScalaCheck in Scala, allowing us to assess the effectiveness of our
technique across different languages and runtime
environments.

This section presents our experimental setup and addresses the following research questions:

\begin{itemize}
  \item \textbf{RQ1}: Do generators written using our technique run faster than those written with regular generator combinators?
  \item \textbf{RQ2}: Do observed generation speedups translate to better bug-finding speed?
  % \item \textbf{RQ3}: Does our technique generalize to other strict functional languages?
\end{itemize}

All experiments were run on a 64-bit Linux machine with 264 GB RAM and a 128-core Intel Xeon Platinum 8375C CPU, running Ubuntu 24.04.1 LTS. \camlname and all OCaml benchmarks
were compiled with 4.14.1+BER MetaOCaml \texttt{ocamlopt}, the native code compiler for OCaml, using compiler flag \texttt{-O3}.
The baseline OCaml generators were written with \bq{} 0.16.
\scalaname{} and all Scala benchmarks
were run on Scala 3.6.3
and OpenJDK 21.0.6, using ScalaCheck version 1.17 as the baseline.
We used \texttt{core\_bench} 0.16~\cite{corebench} in OCaml and \texttt{jmh} 0.4.7~\cite{jmh} for performance microbenchmarking. For assessing and comparing PBT techniques, we used the Etna platform~\cite{etna}.

\subsection{Benchmarking Generator Speed}
\label{subsection:benchmarking}
To answer \textbf{RQ1}, we microbenchmark generators, comparing generators written in \camlname to semantically identical ones in \bq
and generators in \scalaname to those in ScalaCheck.
 We vary the choice of \rand in our \camlname generators, using both \bq's default
\texttt{splittable\_random} and \csplitmix, as discussed
in \secref{subsection:faster-rng}. Our test cases consist of generators for boolean lists, binary search trees (BSTs), and simply
typed lambda-calculus (STLC) terms. We implement generators for these benchmarks using a
variety of \textit{strategies}, varying in structure and
sophistication.

In \camlname, our strategies include: type-derived generators for BSTs and STLC terms,
following the approach described in \secref{subsection:type-derived}; two custom
BST generators---one that builds a tree incrementally by repeatedly inserting
values into an initially empty structure, and another that constructs the
entire tree in a single pass by generating keys, values, and subtrees at each step;
a boolean list generator that mirrors the single-pass BST strategy; and a
generator for STLC terms that is correct by construction (i.e., it produces
only well-typed terms). These microbenchmarks established in the literature\cite{etna} as reasonable proxies for
real-world generators PBT users write.

% These strategies differ in their use of generator library combinators
% and in how frequently they sample random values. The type-derived and
% insertion-based BST generators sample frequently but use combinators
% sparingly. The single-pass BST generator also samples frequently
% but uses more combinators, striking a middle ground.
% The type-derived
% STLC generator samples infrequently and makes moderate
% use of combinators, while the custom STLC generator uses
% significantly more combinators than any other strategy.
% The boolean list generator
% makes moderate use of combinators
% and samples booleans frequently---however, because booleans are
% inexpensive to sample, its overall sampling overhead is lower
% than that of the other strategies.

For each strategy, we compare three treatments: our \bq baseline;
a \camlname version using \bq's randomness
library, \texttt{splittable\_random}; and a
\camlname version using \csplitmix. For each treatment, we measure the
time to generate a value (i.e., a BST, STLC term, or boolean list), using
a random seed, varying generation sizes (10, 100, 1,000, 10,000).\footnote{The
specific meaning of generation size is a domain-specific implementation
detail---in a list generator, size might correspond to the desired length of the list,
whereas in a tree generator it refer to number of nodes, number of leaves, depth of tree,
etc. Regardless, our goal is to show that performance trends scale.}
We run each treatment for 5 seconds and compute the average generation time of a value produced in that interval. Compilation
times are not included, but they are all under 55ms amortized over runs that last on the order of minutes.

\begin{figure}[h]
\ifsilly  \includegraphics[scale=0.66]{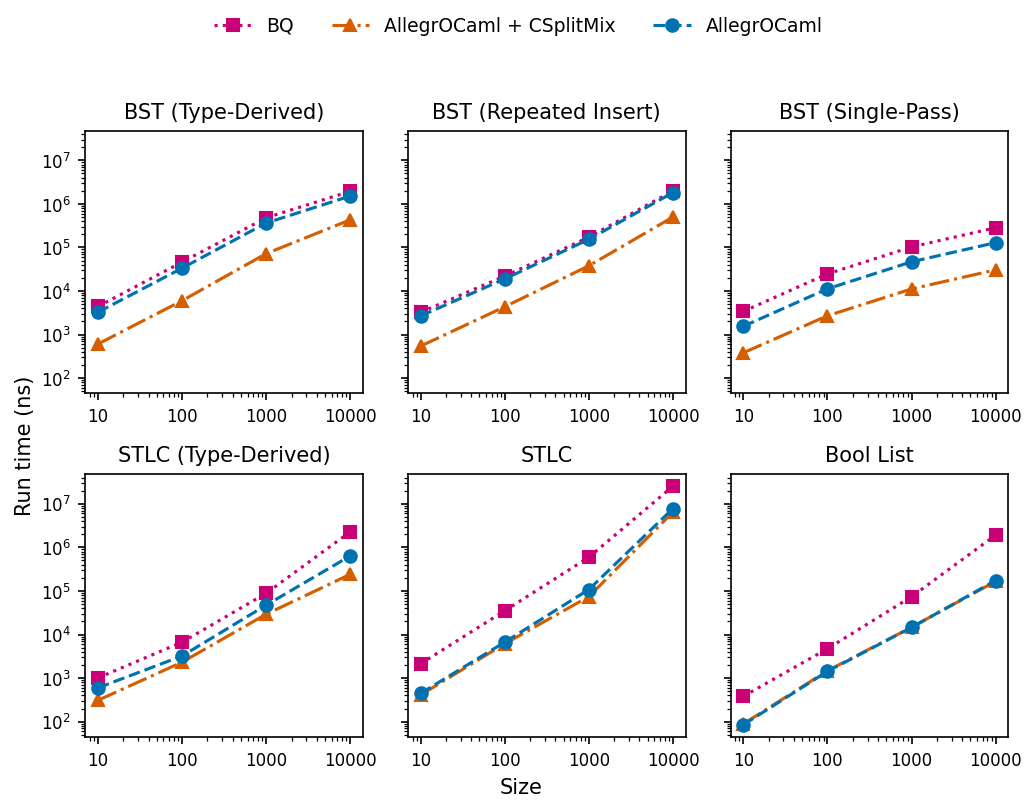} \fi
  \caption{Time to generate values of varying sizes using each \camlname strategy. Lower is better. Both axes are logarithmic. BQ is \bq.}
  \label{fig:time}
\end{figure}

Our results are summarized in \figref{fig:time}.
The type-derived \camlname BST generator achieves speedups ranging from $1.29-1.38\times$, which increase
dramatically, to $4.51-7.70\times$, when combined with \csplitmix. The insertion-based BST generator sees
$1.10-1.21\times$ speedups with staging alone and $4-5.97\times$ when also
using \csplitmix. The single-pass BST generator benefits more from
staging, with speedups of $2.19-2.2\times$, rising to $8.94-9.29\times$ when combined with
\csplitmix.
% We attribute the difference in performance speedup to the fact that
% BST strategies rely on relatively few calls to generator combinator functions,
% but sample extensively from the \rand; notably, single-pass BST, which uses the largest
% number of calls to generator combinators, also sees the greatest benefit from staging. \tr{Call this out earlier? Unless we have the fancy plot.}

For STLC, staging accounts for larger performance gains, yielding
$1.7-3.58\times$ speedups for the type-derived STLC generator and $3.26-5.51\times$ for
the well-typed generator. Adding \csplitmix, these numbers increase to
$2.87-9.46\times$ and $3.94-8.01\times$, respectively.

% This result makes sense because
% STLC generators, particularly the well-typed generator, sample very little from
% randomness libraries but rely extensively on generator library functions.

The boolean list generator experiences $3.3-11.02\times$ speedups with
staging, and its performance changes only minimally when
combined with \csplitmix ($3.2-11.52\times$). This is likely because sampling
booleans is cheap enough that the overhead of crossing the FFI barrier is
comparable to that of generating values directly in OCaml.

The variation in speedups raises a natural question: what
determines how much a generator benefits from staging, \rand optimizations, or both?
One likely explanation is that the two techniques address different performance bottlenecks:
generators that sample more often benefit more from improved random sampling, while those that rely
heavily on generator library combinators gain more from staging.
To test this idea, we use measurable proxies that approximate a generator's reliance on its randomness and generator libraries.
For the former, we count the number of times a generator invokes the SplitMix sampling routine \texttt{next\_int64}.
For the latter, we use the number of calls to \texttt{bind}---the central monadic generator function, extensively used both directly and within other combinators.

We select four strategies with disparate speedup profiles: insertion-based BSTs, single-pass BSTs,
boolean lists, and well-typed STLC terms. To determine how often \texttt{bind} is invoked, we generate 10,000 values using a fixed generation size of 100 and
record the average number of \texttt{bind} calls. To determine the number of random samples, we repeat the same test but use Intel ProcessorTrace~\cite{intelpt} to capture a 4ms snapshot of processor activity.
We then count the number of invocations of \texttt{next\_int64} and average this over the number of generator calls in the trace.

\begin{figure}[h]
\ifsilly  \includegraphics[scale=0.44]{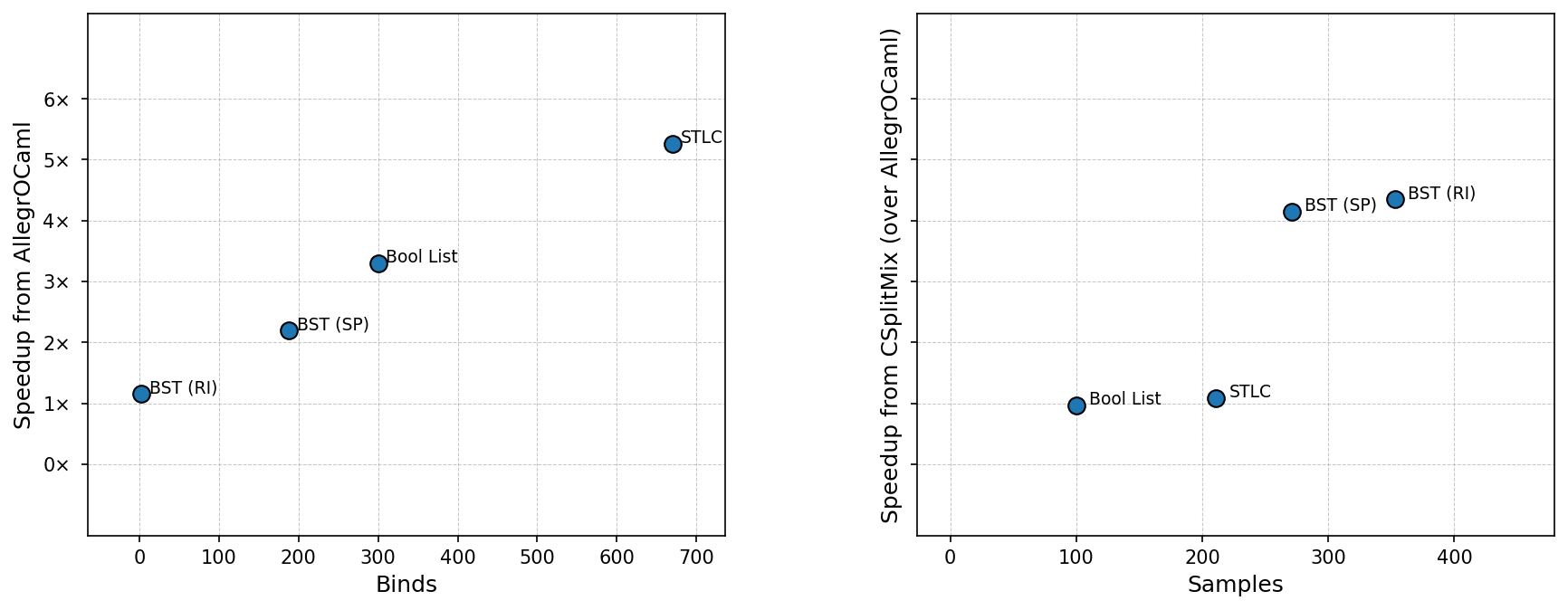} \fi
  \caption{Left: Speedup from staging (compared to \bq) versus the number of \texttt{bind} calls.
  Right: Speedup from \csplitmix (compared to \camlname alone) versus the number of random samples.}
  \label{fig:binds}
\end{figure}

\figref{fig:binds} summarizes the results. The left plot shows a clear linear
relationship between performance benefit from staging and number of calls
to \texttt{bind}. The right plot shows a separation between
generators that sample heavily (BST strategies) and those that
do not (STLC and Bool List), with the former seeing significantly
greater speedups from \csplitmix.

From these results, we conclude that staging and \rand choice play distinct and
complementary roles in generator performance. Sampling-heavy generators see
greater improvements from faster randomness libraries, while combinator-heavy
ones benefit more from staging. Since both factors influence performance
significantly, generator libraries should use both in order to handle diverse
workloads. Further, these results indicate the more complex and sampling-heavy a generator,
the greater \name's benefit. We expect that these results would continue to scale for
even more complicated generators in real-world applications.

In \scalaname, we implement a subset of the \camlname
strategies: the boolean list generator, single-pass BST generator, and
well-typed STLC term generator.\footnote{The type-derived strategies are excluded, since
\scalaname does not currently support type derivation.} We did not implement an optimized version of ScalaCheck's
\rand, but our experimental setup was otherwise the same as in OCaml.

\begin{figure}[h]
  \centering
\ifsilly  \includegraphics[scale=0.68]{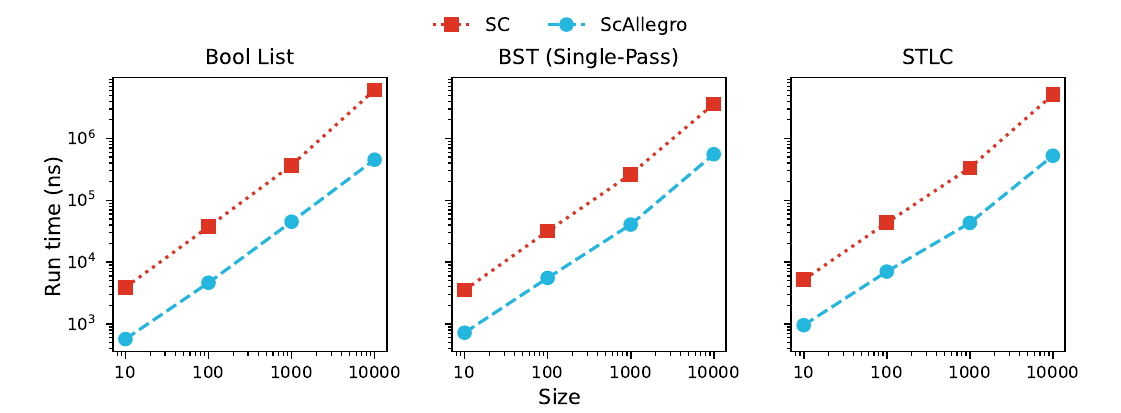}\fi
 \caption{Time to generate values of varying sizes using each \scalaname strategy. Lower is better. Both axes are logarithmic. SC is ScalaCheck; Staged is \scalaname.}
 \label{fig:scala}
 \end{figure}

As shown in \figref{fig:scala}, \scalaname achieves an even greater performance
gain---purely from staging---over ScalaCheck than \camlname does over \bq. In particular,
the single-pass BST strategy is $4.89-6.51\times$ faster, the boolean list generator is $6.86-13.41\times$ faster,
and STLC is $5.43-9.70\times$ faster. This difference arises from
ScalaCheck's representation of generators as functions of type \texttt{size ->
seed -> Option[A]}: each generator combinator must construct and then
pattern-match on these \texttt{Option} values, introducing significant boxing
and unboxing overhead at each step.  Staging eliminates this
overhead. These results show that the \name approach generalizes well across languages.

% \subsubsection{Scala} To show that the benefit of staging extends to
% other strict functional languages, we apply it to ScalaCheck, a widely-used PBT
% library written in Scala.

%\jwc{NOTE: we should test generator speed across both languages, but speed -> bugfinding ability in only OCaml.}
%\jwc{
%  Baseline generators to test speed in both languages:
%  \begin{itemize}
%    \item Single int
%    \item Pair of ints, constrained
%    \item List of ints without bind (use map): lots of sampling, minimal binds.
%    \item Unableled Trees of a fixed size (no weighted union) minimal sampling, lots of binds.
%  \end{itemize}
%}
\subsection{Benchmarking Bug-Finding Speed}
% \bcp{Make sure that capitalization of section titles is consistent}
To answer \textbf{RQ2}, we evaluate the bug-finding speed of our BST and STLC
case studies using Etna~\cite{etna}.
Etna allows users to measure the effectiveness of different generator
implementations by injecting bugs into the system under test and recording the
time taken for a relevant property to fail in response. Each case study includes a diverse
set of \textit{tasks}, where a task consists of a specific bug-property pair
designed to test a specific aspect of the system (e.g., BST includes tasks that
test insertion, deletion, and union operations).
In particular, BST has 37 tasks, and STLC has 20. Strategies for BST and STLC
are implemented in
\camlname; they are unchanged from their description in \secref{subsection:benchmarking}.
Etna does not support Scala, so we were unable to test \scalaname's bug-finding speed in Etna, but we
expect the results in this section to extend to \scalaname.

It is worth noting that the time-to-failure of a given strategy on a given seed
is not necessarily representative of the strategy's average time-to-failure over
a large number of trials. We normalize by computing the relative performance, or
the ``speedup.'' This works in most cases, but it is not perfect.  For example,
it is theoretically possible to choose a seed such that the \textit{first} value
produced by a generator discovers the bug, which would eliminate any differential
speedup, and likewise if both generators fail to find the bug.  To account for
these cases, we repeat the above process over 30 random seeds---that is, all strategies
run on the same seed so that they produce identical sequences of values, and this process
is repeated 30 times. Although the
variance \textit{between} seeds in Etna can be vast, timing results are
replicable for a \textit{given} seed. Across 1,000 trials using the same seed,
the observed variance in time-to-failure was less than a nanosecond.

We run each
strategy on all tasks using a 60-second timeout.  If a bug is not found within
this limit, the task is considered ``unsolved.''  We exclude tasks where all strategies fail to
find the bug from our dataset, as they provide no basis for
comparison.  Similarly, we exclude tasks where the \bq strategy completes in
under 5ms, as such negligible run times do not yield meaningful insights into
relative performance. Across all 30 seeds, these filters remove
28/600 (4.67\%) of type-derived STLC's tasks,
168/600 (28\%) of STLC's tasks,
149/1110 (14.42\%) of type-derived BST's tasks,
268/1110 (24.1\%) of single-pass BST's tasks, and
371/1110 (33.42\%) of insertion-based BST's tasks. More sophisticated
strategies tend to find bugs very quickly, leading to a higher number of filtered
tasks. By applying these filters, we ensure that our reported speedups reflect
optimizations that meaningfully impact performance.

\begin{figure}[h]
\ifsilly  \includegraphics[scale=0.59]{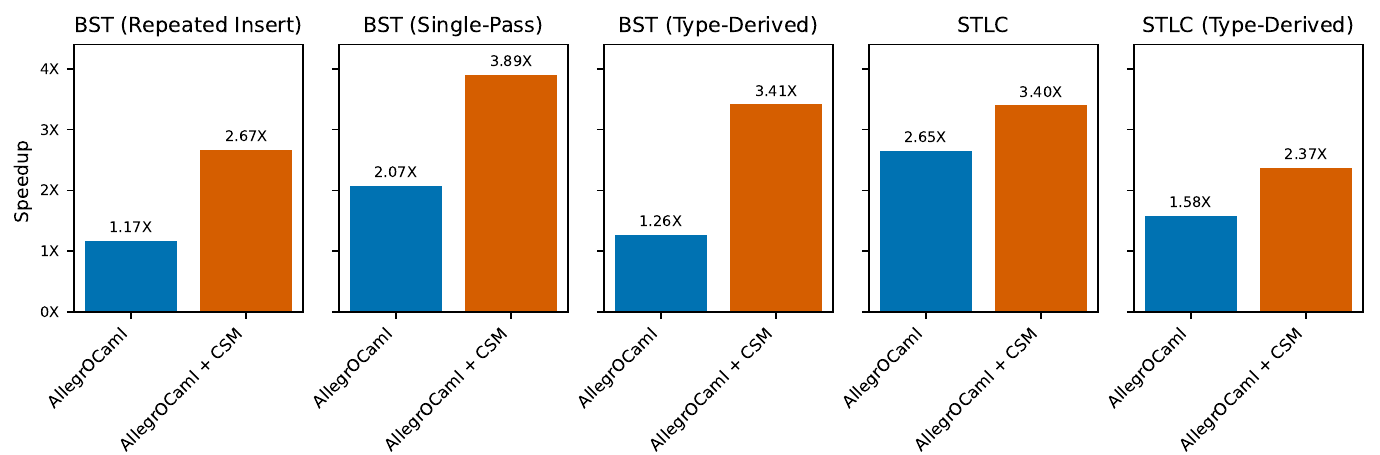}\fi
  \caption{Geometric average of all speedups---relative to \bq---for each strategy and benchmark, showing that staging leads to better bug-finding speed across the board. CSM is \csplitmix.
  }
  \label{fig:etna}
\end{figure}

\begin{figure}[h]
\ifsilly  \includegraphics[scale=0.36]{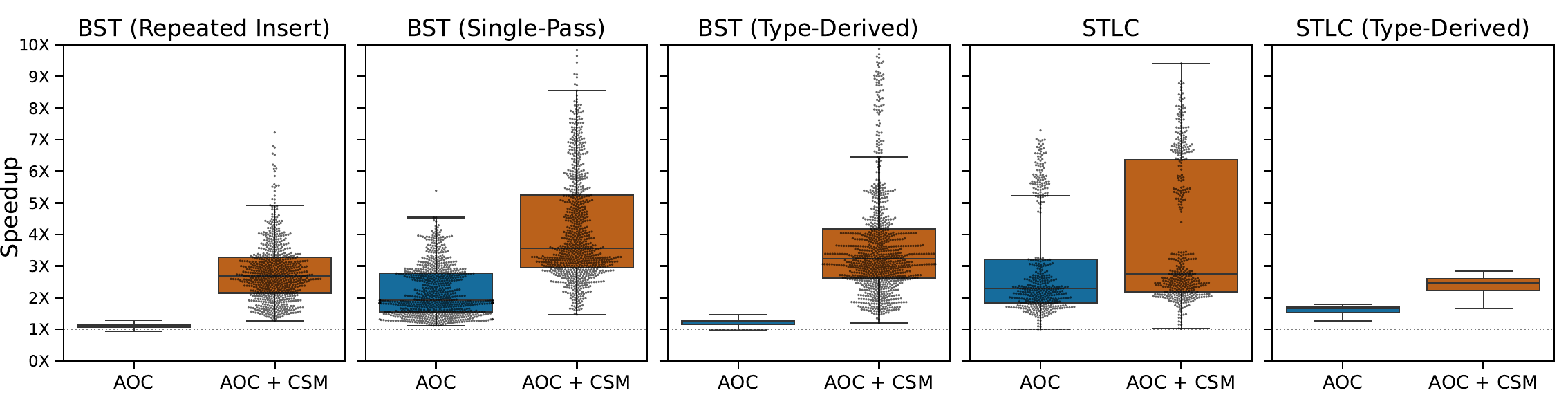} \fi
  \caption{Distribution of individual-task speedups across strategies and benchmarks. Gray dots represent tasks. Plots with extremely low variance have no swarm overlay. CSM is \csplitmix; AOC is \camlname.}
  \label{fig:swarm}
\end{figure}

Our results are summarized in \figref{fig:etna}, which shows the geometric average of
individual-task speedups for each strategy and benchmark. Trends in bug-finding speed reflect trends in performance
from \secref{subsection:benchmarking}. The distribution of speedups in \figref{fig:swarm} reveals
substantial variability, with most tasks clustering near the median and a long tail of outliers
achieving much larger gains. STLC shows a more bimodal distribution of speedups
than the other benchmarks, which we attribute to the heterogenous difficulty of its tasks:
some tasks regularly hit the 60-second timeout, while others finish in a fraction of a second. We
find that ``easier'' tasks---those that run on the order of milliseconds---regularly achieve speedups in the
range of $4-9\times$, whereas tasks that run on the order of seconds achieve more moderate speedups of up to $3.5\times$.
Overall,
these results give strong evidence that \camlname consistently benefits bug-finding speed, sometimes drastically. We expect that these results would
hold in real-world scenarios where generation time constitutes a significant proportion of total testing time.

In addition to \camlname's overall bug-finding speed, we evaluate
the performance of its type-derived generators---specifically,
whether they can transform previously intractable tasks into
tractable ones. Out of 79 tasks that initially timed out,
31 (38.4\%) completed successfully using \camlname alone, and
51 (64.1\%) completed successfully after adding \csplitmix.
This shows that staged type-derived generators can provide
significant performance improvements for "free"---that is, without
 any additional effort from the user.
 
\section{Other Languages \& Libraries}
\label{section:other-langs}

In the years since the original QuickCheck paper~\cite{quickcheck00}, PBT has had remarkable success
in languages outside of the Haskell world from which it came~\cite{bq,sc,hypothesis}.
For this reason, readers who are users and developers of PBT libraries in \emph{other} languages---including non-strict or procedural ones---might be curious about how the
techniques brought to bear here on OCaml's \bq{} and Scala's ScalaCheck
might be imported to their favorite language.

We begin by noting that the principle of choosing fast random number generators
is completely language-agnostic. In languages like OCaml, where run-time value representations
make natively-implemented \rand inefficient, calling out to a C implementation or using standard libraries written in C is a surefire win.
In other languages, serious thought should be put into using as fast of a \rand{} as possible,
as opposed to picking up any suitable option off the shelf.
Next, we discuss language-by-language the degree to which the staging-related insights of this paper are portable.

  \textbf{Racket}'s entire philosophy is intertwined with
  using macros to build small eDSLs like the ones we use to write generators, and so it seems a natural fit.
  However, the Racket macros literature, while extensive, does not usually concern itself with staging for performance purposes.
  For this reason, we---the authorship team of this paper, devoid of much Racket expertise---decided to not use Racket as a test case in this paper.

  \textbf{Haskell} is home to the original PBT implementation, and
  so it is natural to ask why we have not applied \name{} to it yet. The basic
  answer is that the Haskell compiler (GHC) is designed specially to eliminate
  the run-time overhead of monadic abstractions! Indeed, because Haskell is
  pure, GHC can aggressively inline and beta-reduce nearly anything as part of
  its normal optimiation steps.  This means that in many cases, the impact of a
  staged PBT library (and indeed, all staged monadic DSLs) is negligible.  For
  complex enough programs, however, the heuristics GHC uses can degrade, leading
  to performance overheads for monadic code.  For this reason, Haskell does have
  a multi-stage programming system, Template Haskell~\cite{templatehaskell}, that is
  sometimes used to build eDSLs in contexts where one does not want to rely on the
  optimizer's heuristics.  In short: it is possible to build an \name{} version of Haskell
  QuickCheck, but the benefits would be much less pronounced.

  \textbf{F\#} has both multi-stage programming capabilities~\cite{fsmeta} and a well-used PBT library called FsCheck~\cite{fscheck}.
  The internals of FsCheck's generator eDSL are very similar to both
  \bq and ScalaCheck, so we expect that building \name{}-style generators
  in F\# should be a straightforward engineering effort.
  % https://learn.microsoft.com/en-us/dotnet/fsharp/language-reference/code-quotations

  \textbf{Rust} shares some similarities with functional languages, and,
  indeed, it features a number of property-based testing libraries~\cite{Bolero,ProptestBook}. However, Rust does not
  structurally encourage monadic eDSLs---it does not have a special monadic syntax---and so
  PBT libraries in Rust ask programmers to define generators directly as \texttt{seed -> value} functions.
  For this reason, staging does not seem directly applicable to any of
  the PBT libraries in Rust.

  \textbf{Clojure}'s PBT library test.check\cite{test.check} is a straightforward port
  of Haskell's QuickCheck. Like Racket, 
  Clojure has a powerful macro system which could be employed to build staged generators. Further, since
  Clojure runs on the JVM, we anticipate that the performance benefits of staging
  would be similar to those observed in ScalaCheck.

  \textbf{Swift}'s PBT library SwiftCheck\cite{swiftcheck} is likewise a direct port of Haskell's QuickCheck,
  and Swift's macro system\cite{swiftmac} is expressive enough to support staged generators. Swift's inlining is
  not as robust as GHC's, so we expect that it would also see some benefit from the \name{} technique. However,
  SwiftCheck is not maintained, so we did not pursue this direction.

This covers a large swath of functional and functional-adjacent languages with both PBT libraries and metaprogramming capabilities,
suggesting that the Allegro technique is widely applicable.

\section{Related Work}
\label{section:related}

\paragraph{Speeding up Property-Based Testing}

Our work is unique in its approach to speeding up
PBT, but it is certainly not the only work focused on making PBT faster.

Changing the shape of the test input distribution is perhaps the most well-studied way to accelerate bug-finding in PBT.
 Feedback-driven mechanisms
like Targeted PBT~\cite{loscherTargetedPropertybasedTesting2017} and
Coverage-Guided PBT~\cite{lampropoulosCoverageGuidedProperty2019}
speed up testing by changing the generation order to find
``interesting'' inputs faster. Separately, enumerative approaches to
PBT~\cite{runcimanSmallcheckLazySmallcheck2008,braquehaisSimpleIncrementalDevelopment2017}
try to get to bugs faster by leveraging the ``small-scope
hypothesis,'' which posits that most bugs will be triggered by smallish
inputs. All of these approaches can have significant performance
benefits in
practice, but they are largely orthogonal to our contributions; we expect
both \name{} and faster \rands{} could be used to speed these existing techniques to varying
degrees.

Similarly orthogonal to our approach are techniques for quickly filtering out
invalid inputs. Some of these approaches work statically,
by automatically deriving generators that produce only valid inputs by
construction~\cite{lampropoulosGeneratingGoodGenerators2017}, while others
filter dynamically via laziness~\cite{claessenGeneratingConstrainedRandom2015}
or by solving satisfiability
problems~\cite{seidelTypeTargetedTesting2015,steinhofelInputInvariants2022,lampropoulosBeginnerLuckLanguage2017}.
In all cases, these approaches could be further improved using
techniques from \name and faster \rands{}.

The most direct comparison to our work is
QuickerCheck~\cite{krookQuickerCheckImplementingEvaluating2024}, an
implementation of Haskell's QuickCheck that exploits the inherent parallelism of
PBT to achieve significant performance gains. QuickerCheck, like \name, was
designed with performance engineering in mind, taking seriously the idea that
PBT is a performance critical task. Still, despite their similar
motivations, the solutions are entirely different and complementary.

\paragraph{Multi-Stage Programming}
% \jwc{A generator is a parser of
% randomness~\cite{goldsteinParsingRandomness2022}, and people have implemented
% lots of staged parser libraries!}
Staging's roots come from quasiquotation in LISP \cite{bawden99}, where the
unified representation of data and code allows for sophisticated
metaprogramming. Quasiquotation-style macros were introduced to the ML family of
languages by Nielson and Nielson's Two-Level Functional Languages~\cite{nielson92} and
MetaML~\cite{taha00}, which in turn quickly inspired MetaOCaml~\cite{metaocaml} and more recently
MacoCaml~\cite{macocaml}, Lightweight Modular Staging in Scala~\cite{lms}, Template
Haskell in Haskell~\cite{templatehaskell}, and more.  In addition to implementations, researchers
have studied the type-theoretic foundations of multi-stage languages \cite{cmtt,
davies-pfenning, davies17}, as well as ways of combining multi-stage types with
other sophisticated features like dependent types \cite{kovacs24}. Meanwhile,
the Racket community continues the LISP tradition with a long and fruitful
line of work studying how macros and metaprogramming can be used to build extensible
DSLs \cite[etc.]{flatt12, samth11, skthesis}.

In ML-like languages, the primary use of staging is building embedded DSLs
with minimal performance overhead, an idea that has come to be known as ``abstraction without regret''
\cite{lms,parreaux17,parreauxthesis, sheard99, trattdsl}.
This technique has been applied across a range of domains, including
big-data processing \cite{jet12}, stream processing \cite{moller20,strymonas},
query processing \cite{rhyme}, and parser combinators \cite{sspc, staged-parsers, krishnaswami19, flap}.
This work on staged parsers is the closest analogue to ours---indeed, PBT research has drawn an explicit link between parsing and generation, framing generators
as ``parsers of randomness'' \cite{goldsteinParsingRandomness2022}.
However, this equivalence is primarily theoretical. PBT generators (a)
are not actually implemented as parsers
of random sequences and (b) support choice-based combinators like \texttt{weighted\_union} that parsers do not.

% \paragraph{Sources of Randomness in Random Testing}
% \jwc{
%   \begin{itemize}
%     \item original QC work used SplitMix \cn{}, to split at binds.
%     \item the point of
%   \end{itemize}
% }

\section{Conclusion, Limitations, \& Future Work}
\label{section:concl}
We have identified, studied, and proposed general solutions to two
important sources of inefficiency in PBT generator libraries: abstraction
overhead and choice of \rand{}.

Our approach has limitations. First, it requires the implementor of a generator 
library to write hand-staged versions of its combinators. This can be an 
arduous process, particularly in host languages with poor meta-programming 
support; moreover, users must program against the staged implementation, 
which is a departure from the standard model. Second, the benefit of the 
Allegro technique is inversely tied to the host compiler’s ability to 
eliminate generator library overhead: we observe larger speedups in Scala 
than in OCaml (and smaller speedups in Haskell). Our evaluation used OCaml’s 
standard compiler, whose shorter compile times suit development environments 
where tests are usually run; had we used another version of the compiler 
(e.g., one using flambda\cite{ocaml-flambda}), our experimental results would naturally have 
been different.

In the future, we hope to continue to investigate and push the boundaries of
PBT generator performance.
On the abstraction overhead side, we hope to employ some of the many tricks for better code generation that have been written
about in the staging literature. For example, MetaOCaml includes a primitive floating let-insertion~\cite{kiselyov2022letrecinsertioneffects} that yields optimal let placement,
which we did not use in \camlname{}.
Another trick we hope to use is GADT-based techniques for unpacking the states of recursive functions~\cite{kovacs24}. Currently, the \camlname{}
recursion combinator introduces some overhead from boxing and unboxing the accumulator values at each recursive call. As noticed in other work \cite{kovacs24}, this
overhead can be eliminated using a type-level heterogenous list to represent the state at compile time.

On the randomness side we plan to investigate fast \rands{} that are not equivalent to SplitMix, such as Lehmer~\cite{lehmerrng}, WyHash~\cite{wyrand},
and variants of xoroshiro~\cite{xorshiro}. We also want to try techniques to speed up sampling such as pipelining, using SIMD instructions,
or generating large buffers of random numbers ahead of time. Finally, we hope to investigate the degree to which statistical randomness
matters in PBT, since some of these ideas trade statistical guarantees for speed.
\begin{acks}
This work was partially funded by the U.S. National Science Foundation under Grant No.~2402449. Special thanks to Brett Saiki and Daniel Sainati for their time reviewing our paper.
\end{acks}

\subsection*{Data Availability Statement}
For artifact evaluation we have submitted (1) the source code of \camlname{} and \scalaname{} as well as (2) the code for our version of Etna,
which we have modified slightly to (a) support parallelism, and (b) pass deterministic seeds to all versions of a single strategy.
These programs are publicly available~\cite{artifact}.
The dataset for our evaluation is produced by a combination of Etna and microbenchmarks in the \camlname{} and \scalaname{} sources.

\bibliographystyle{ACM-Reference-Format}
\bibliography{bib,harry}

\end{document}

%% file: macros.tex
\definecolor{dkred}{rgb}{0.7,0,0}
\definecolor{dkblue}{rgb}{0,0,0.7}
\definecolor{notered}{rgb}{0.85,0,0}
\definecolor{dkpurple}{HTML}{4e02eb}
\definecolor{dkgreen}{HTML}{006329}
\definecolor{teal}{HTML}{007982}
\definecolor{fuchsia}{HTML}{8C368C}

\newcommand{\comm}[3]{\ifdraft\textcolor{#1}{[#2: #3]}\fi}

\newcommand{\hg}[1]{\comm{teal}{HG}{#1}}

\newcommand{\bcp}[1]{\comm{dkred}{BCP}{#1}}

\newcommand{\bq}{\texttt{Base\_quickcheck}\xspace}
\newcommand{\name}{{Allegro}\xspace}
\newcommand{\camlname}{AllegrOCaml\xspace}
\newcommand{\scalaname}{ScAllegro\xspace}

\newcommand{\rand}{randomness library\xspace}
\newcommand{\rands}{randomness libraries\xspace}

\newcommand{\secref}[1]{\S\ref{#1}}
\newcommand{\csplitmix}{CSplitMix\xspace}
\newcommand{\figref}[1]{Figure~\ref{#1}}